\documentclass[%
    reprint,
    superscriptaddress,
    longbibliography,
    nofootinbib,
    amsmath,
    amssymb,
    aps,
    floatfix,
]{revtex4-2}
\usepackage{graphicx}
\usepackage{dcolumn}
\usepackage{bm}
\usepackage[utf8]{inputenc}
\usepackage{acronym}
\usepackage{graphicx}
\usepackage{dsfont}
\usepackage{amsmath,amsfonts,amssymb,amsthm}
\usepackage{braket}
\usepackage{physics}
\usepackage{tikz}
\usetikzlibrary{quantikz}
\usepackage{acronym}
\usepackage{xspace}
\usepackage{siunitx}
\usepackage{booktabs}
\usepackage[hidelinks]{hyperref}
\usepackage{nicefrac}       
\usepackage{microtype}      
\usepackage{mathtools}

\usepackage{listings}
\usepackage{todonotes} 
\usepackage{siunitx}
\usepackage{multirow}
\usepackage{tabularx}

\usepackage{booktabs, multirow, siunitx}
\usepackage[table]{xcolor}

\hypersetup{
    colorlinks=true,
    urlcolor=blue,
    citecolor=blue,
    linkcolor=blue
}

\begin{document}

    \title{Characterization and upgrade of a quantum graph neural network for charged particle tracking}
    
    \author{Matteo Argenton}
    \email{matteo.argenton@unife.it}
    \affiliation{Department of Physics and Earth Science, University of Ferrara, Via Saragat 1, Ferrara, 44122, Italy}
    \affiliation{INFN Ferrara, Via Saragat 1, Ferrara, 44122, Italy}
    
    \author{Laura Cappelli}
    \affiliation{INFN Ferrara, Via Saragat 1, Ferrara, 44122, Italy}
    
    \author{Concezio Bozzi}
    \affiliation{INFN Ferrara, Via Saragat 1, Ferrara, 44122, Italy}
    
    \date{\today}

\begin{abstract}
    In the forthcoming years the LHC experiments are going to be upgraded to benefit from the substantial increase of the LHC instantaneous luminosity, which will lead to larger, denser events, and, consequently, greater complexity in reconstructing charged particle tracks, motivating frontier research in new technologies. Quantum machine learning models are being investigated as potential new approaches to high energy physics (HEP) tasks. We characterize and upgrade a quantum graph neural network (QGNN) architecture for charged particle track reconstruction on a simulated high luminosity dataset.
    The model operates on a set of event graphs, each built from the hits generated in tracking detector layers by particles produced in proton collisions, performing a classification of the possible hit connections between adjacent layers. In this approach the QGNN is designed as a hybrid architecture, interleaving classical feedforward networks with parametrized quantum circuits. We characterize the interplay between the classical and quantum components. We report on the principal upgrades to the original design, and present new evidence of improved training behavior, specifically in terms of convergence toward the final trained configuration.
\end{abstract}

\maketitle

\section{Introduction}\label{sec:intro}

    Within high energy physics (HEP), the reconstruction of the trajectories of charged particles, starting from the clustered energy deposits left during their propagation through the tracking layers of a detector, is a computationally intensive, but central task~\cite{CERN-LHCC-2022-005}. \\At the luminosity regimes that the Large Hadron Collider (LHC) is entering in 2030, which marks the beginning of its high luminosity phase (HL-LHC)~\cite{Apollinari:2015wtw}, particle tracking in the two major general purpose experiments, CMS and ATLAS, will need to account for an increment of a factor 3-5 in the average number of primary proton-proton inelastic interactions (referred to as \emph{pileup}), averaging 140–200 in future runs~\cite{Brüning:2906155}. Similar upgrades, although targeting different levels of luminosity, are also planned for the LHCb~\cite{LHCbcollaboration:2903094} and ALICE~\cite{Dainese:2925455} detectors at a later stage. 
    \\While the tracking pipelines currently in use up to Run 3~\cite{CMS_tracking_2014, Scarabotto_HLT1, HLT2_lhcb, ATLAS_run3_tracking} are being upgraded to handle the combinatorial blow-up, which is a consequence of the increased proton flux, through e.g. pileup mitigation techniques~\cite{Bertolini:2014bba, CMS:2020ebo}, these tracking stacks are often rooted in the combinatorial Kalman Filter (CKF) algorithm~\cite{Billoir:1989mh, Magano:2021jzd}, and are therefore inherently challenged because of their local approach to track reconstruction.
    
    In recent years, a plethora of novel approaches has emerged, mostly based on machine learning models, and targeting different stages of the reconstruction pipeline. In particular, graph neural networks (GNN) are one of the most promising technologies, both specifically for particle tracking~\cite{ExaTrkX:2021abe, Acorn:2024, Correia:2919388}, and for a broader class of particle physics applications~\cite{Duperrin:2023elp, Qasim:2022rww, Aurisano:2024uvd}.
    
    Alongside these ongoing developments in machine learning for high energy physics, other novel computing paradigms are being researched. In particular, quantum technologies are explored as alternative, and often synergic, computing paradigms for solving classically inefficient tasks.
    \\Within this scope, quantum machine learning (QML) is a hybrid field targeting to understand how learning from data can be performed through quantum algorithms, and it has seen a considerable development in the last decade, often in parallel with, but not constrained to, modern machine learning models.
    \\The research presented in this work focuses on a QML application to the specific task of track segment classification in particle tracking.  
    We discuss the development, characterization, and upgrade of a quantum graph neural network (QGNN) model first proposed in~\cite{tuysuz2021hybrid}, and here extended to provide improved training and generalization performance.
    \\QGNNs belong to the broad class of variational quantum machine learning algorithms that can be tested in the current NISQ (Noisy Intermediate Scale Quantum) era, where the number of qubits and circuit depth are limited both for quantum hardware, and classical simulators. In particular, QGNNs are an advanced application of parametrized quantum circuits (PQC), to classical graph neural networks. This class of architectures exploits the flexibility and representation capacity of classical multi-layer perceptrons (MLP), to feed and extract data from the circuits.
    \\This study is aimed at improving the understanding of hybrid variational quantum algorithms, and their possible applications in experimental particle physics tasks.

\section{Dataset and event graph construction}

    The high pileup data used in this research comes from the simulated TrackML dataset\cite{trackml-particle-identification}, which is a standard reference for machine learning research on particle tracking. The dataset is based on a modular, synthetic tracking detector whose design is based on the high-luminosity upgrades of both the CMS and ATLAS inner trackers; additional details on the detector are reported in Appendix~\ref{A:TrackML}. 

    Before being used as input to the QGNN, the TrackML events undergo a preprocessing step, consisting of a set of constraints on the hits' physical coordinates -- necessary to reduce the number of hits and, consequently, of candidate track segments -- and a graph-building stage, where an input graph is built from the selected hits of each collision event.

    \subsection{Preprocessing constraints}

\begin{figure}[t]
        \centering
            \includegraphics[width=1\linewidth]{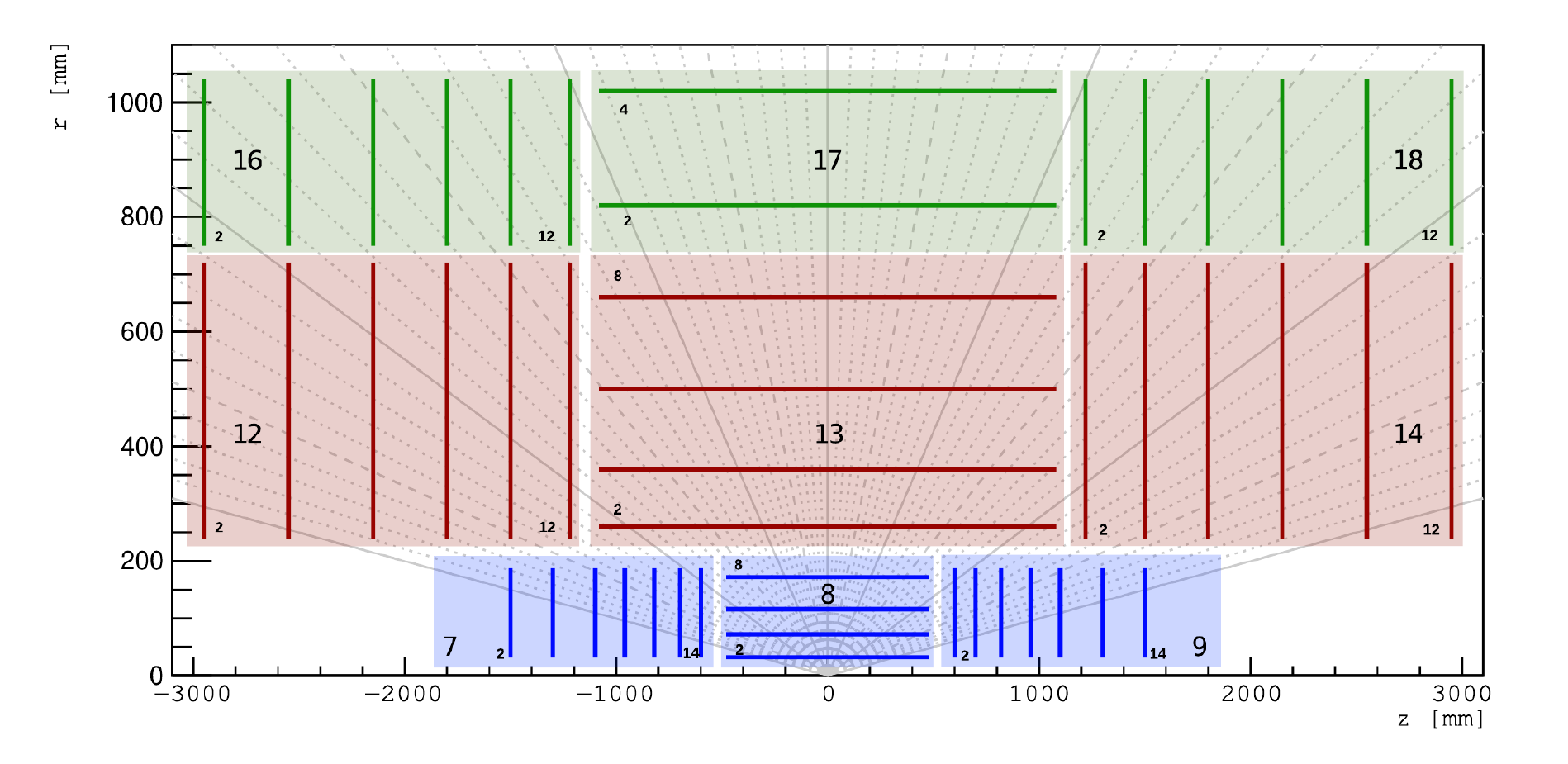}
            \caption{TrackML detector layout from~\cite{trackml}. The three different sub-detectors are shown on the left and with the associated colors in the z-r plane depiction on the right. Grey lines correspond to different pseudorapidity $\eta$ values.}
            \label{fig:TrackML_detector}
    \end{figure}

        In the tests presented in this work only the hits in the central (barrel) regions of Fig.~\ref{fig:TrackML_detector} are considered; for a total of ten tracking planes.
        This restriction can be lifted in future studies, but represents a good compromise, especially in the hybrid quantum framework, between the time needed to train the models, and the fidelity to a real-world tracking scenario.
    
        Similarly, a cut is enforced on the particles' transverse momentum, to only retain $p_T\ge 1$GeV tracks. This cut is also motivated by the training time of the hybrid quantum model, both when the quantum resources are simulated on classical hardware, and when tests are performed on quantum hardware.
        
        In this research the geometrical constraints on the hits are aligned with the ones of the original QML study~\cite{tuysuz2021hybrid}, and of the classical HEP.TrkX project~\cite{farrell2018noveldeeplearningmethods} from which it originates. The coordinates of each hit are expressed in a cylindrical reference system $(r,\phi,z)$, where $r$ is the radial distance from the beam axis $z$, and $\phi$ is the azimuthal angle. For each hit pair $(h_1,h_2)$ the following quantities are computed and constrained:
        \begin{itemize}
            \item $\phi_{\text{slope}}=\Delta \phi / \Delta r$
            \item $z_0 = z_1-r_1 \Delta z / \Delta r$
        \end{itemize}
        with cuts:
        \begin{itemize}
            \item $|\phi_{\text{slope}}|<6 \times 10^{-4}$
            \item $\left|z_0\right|<100 \mathrm{~mm}$.
        \end{itemize}

    \subsection{Graph building}

        A graph $\mathcal{G}=(\mathcal{V}, \mathcal{E})$ is a collection of objects $\mathcal{V}$ (named nodes, or vertices), together with a set of relations between said objects $\mathcal{E}$ (edges, or arcs). Given two nodes $i, j \in \mathcal{V}$; the edge going from $i$ to $j$ is $e=(i,j) \in \mathcal{E}$. If the edge has no preferential direction, hence $(i,j)\equiv(j,i)$ the edge is said to be undirected; if the opposite is true it is a directed edge. This work considers only directed edges because they have shown to consistently outperform models of the latter kind.
        \newline
        Edges can be both weighted or unweighted. An arbitrary graph is weighted if it is equipped with a weight function \(w:\mathcal{E}\to\mathbb{R}\), where the weights $w_e$ are all non trivially one.
    
        In each collision event from the dataset, a graph edge connecting two hits is generated whenever two hits belonging to adjacent layers are geometrically compatible. The resulting structure is indeed a graph, where the edges are candidate track segments, and the nodes are hits.
    
        The geometrical constraints from the previous section are particularly important in this construction as they guarantee that each graph --with average number of nodes $\bar{N}_{\mathcal{V}}=(5.1\pm 0.8)\cdot 10^3$ and number of edges $\bar{N}_{\mathcal{E}}=(7.7\pm 1.7)\cdot 10^3$-- has on average $54\pm4 \%$ of true track segments. This makes the dataset balanced.

        The connectivity of the graphs built according to this procedure are described through the \emph{incident matrices}, here defined according to the conventions of~\cite{tuysuz2021hybrid}:
        \begin{equation}
        \label{R_i}
            R_i^{j k}=\left\{\begin{array}{l}
            1, \text { if } k^{t h} \text { edge is input of } j^{t h} \text { node } \\
            0, \text { otherwise }
            \end{array}\right.
        \end{equation}
        \begin{equation}
        \label{R_o}
            R_o^{j k}=\left\{\begin{array}{l}
            1, \text { if } k^{t h} \text { edge is output of } j^{t h} \text { node } \\
            0, \text { otherwise }
            \end{array}\right.
        \end{equation}
        where to define the $k^{th}$ edge it is assumed that the set of edges $\mathcal{E}$ of the event graph $\mathcal{G}$ is ordered.
        These matrices in Eq.\eqref{R_i} and Eq.\eqref{R_o} are used to, respectively, address the incoming and outgoing edges, given a node $j$.

\section{Graph Neural Networks}

        Among the different intersections of graph theory and machine learning, graph neural networks represent one of the most successful deep learning models of the past decade~\cite{wu2020comprehensive}. GNNs are born from the idea of generalizing the convolutions, performed by a convolutional neural network in euclidean domain (e.g. the ones on a 2D pixel grid), to the non-euclidean domain of a graph~\cite{bronstein2017geometric}.
        The fundamental idea is that, in graph domain, notions like shift invariance find no obvious counterpart and have to be adapted in a meaningful way from first principles.

        Learning on graphs as a concept has first been introduced in the context of recurrent neural networks and random walk models in~\cite{gori2005new}. Subsequent studies have appeared much later, and generally come from two different, but intertwined approaches to graph-like domains: on one side works explicitly based on spectral graph theory~\cite{bruna2013spectral, chebnet, kipf2016semi}, and on the other research considering graphs as a mesh, and therefore rooted in differential geometry\cite{masci2015geodesic, boscaini2016learning, monti2017geometric}.
        Modern approaches~\cite{velivckovic2017graph} draw inspiration from both these frameworks. In particular the classical computing architecture used in this project is deeply inspired by Exa.TrkX project~\cite{ExaTrkX:2021abe}, and, specifically, its attention-based GNN.
        The two most important features in this class of GNNs are (i) message passing, and (ii) edge attention.

        \begin{figure}
        \centering
            \includegraphics[width=1\linewidth]{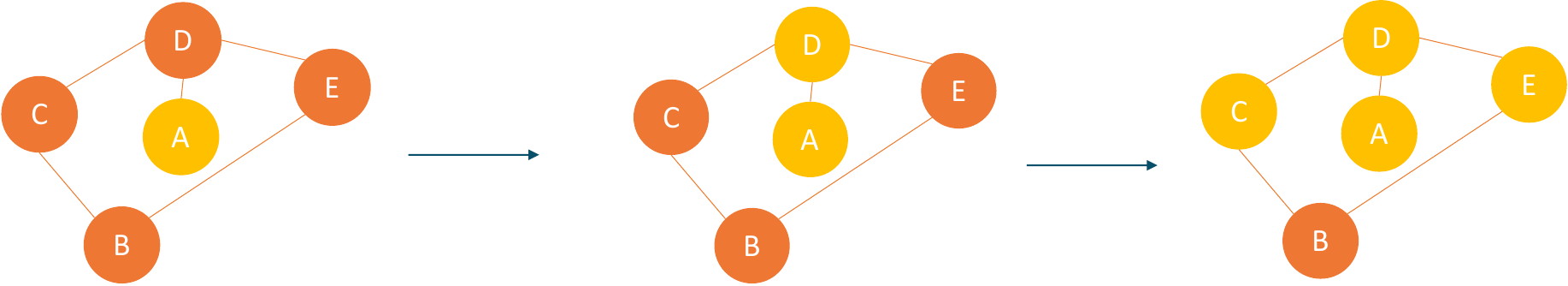}
            \caption{Message passing algorithm. Information from node A is first propagated to its nearest neighbor D, and then propagated to the next to nearest neighbors C and E.}
            \label{fig:Node messa}
        \end{figure}
        
        Message passing refers to the propagation of information within the graph, and is illustrated in a simplified scenario in Fig.~\ref{fig:Node messa}. Each node in the graph iteratively enriches the feature vector that describes its properties, aggregating (for example concatenating) information coming from its neighbors. This is the generalization of the convolution operation that a CNN performs on a regular grid, but on a graph domain, where the connectivity is not necessarily regular.
        \newline
        Every incoming message is weighted by a $[0,1]$ attention score, which is a scalar weight associated to each edge. 
        This double-sided mechanism guarantees that the nodes that in the initial graph are connected with each other --comprising both true and fake candidate track segments-- at multiple levels of proximity (different detector layers in the case of graphs representing collision events), become aware of their neighborhood.
        
        For these reasons GNNs are an example of a modern, machine-learning-based, global method to particle track building.
        \newline
        Information propagation happens for each node (and edge) at the same time step, within a single global environment, rather than building the track one node at a time for each separate instance of a combinatorial Kalman filter (refer to~\cite{CMS_tracking_2014, Magano:2021jzd} for a full description of such a pipeline). It is worth highlighting that a full, CKF-based track reconstruction pipeline comprises many steps, from seeding, to finding, fitting, and selection; these steps are not performed by a single GNN. Modern GNN-based projects for tracking in HEP are multi-stage frameworks~\cite{ExaTrkX:2021abe, Acorn:2024}, where the central part is a track segment classification performed by a graph neural network. The output of this stage is a high-accuracy set of true track segments between pairs of hits from adjacent layers (doublets). These doublets are then further aggregated in triplets (three connected hits), and, finally, in full tracks, by the subsequent routines in the tracking pipeline.
        
        Similarly to these classical pipelines, the quantum machine learning model presented in this work targets the track segment classification stage.
        
        \subsection{Structure of a GNN for edge classification}

            \begin{figure*}
            \centering
                \includegraphics[width=0.7\linewidth]{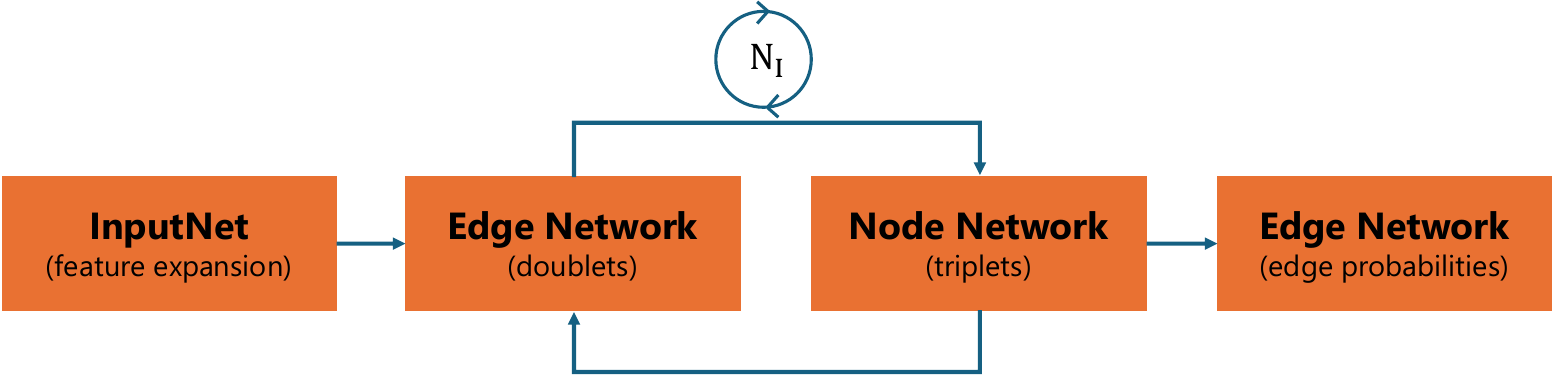}
                \caption{Block diagram of a graph neural network for edge classification. The iteration of Edge and Note Networks is responsible for the propagation of the information at different orders of proximity in the graph structure, as in Fig.~\ref{fig:Node messa}.}
                \label{fig:gnn}
            \end{figure*}

            The message passing and attention methods described in the previous section are implemented at different stages within the GNN, and are illustrated schematically in Fig.~\ref{fig:gnn}. The iteration between the Edge and Node Network blocks, which are individually discussed in Appendix~\ref{A:GNN}, is the mechanism responsible for the propagation of the information at an increasingly distant number of hops in the event graphs, and for this reason it complements the message passing executed within the Edge and Node Network blocks.

            \subsubsection{InputNet}
                A graph built according to the procedure detailed in the preprocessing section can be represented through three matrices. The feature matrix $X\in\mathbb{R}^{N_{\mathcal{V}}\times3}$, where each row is a tuple consisting of three hit coordinates, and the incident matrices $R_i,\ R_o\in {\{0,1\}}^{N_{\mathcal{V}}\times N_{\mathcal{E}}}$ introduced in Eq.\eqref{R_i} and Eq.\eqref{R_o}. A final ground truth vector $y\in{\{0,1\}}^{N_{\mathcal{E}}}$ completes the set of preprocessed outputs.

                The input features of the hits of a graph are fed to the InputNet node-by-node for each event. Each step of the GNN acts on all the nodes (or edges) before moving to the following stage. In this sense the InputNet is a MLP which, for each row in $X$, expands its representation to express correlations between the three physical input coordinates.

            \subsubsection{Edge Network}
                The first and last stage of the GNN are Edge Networks (EN). All ENs share the same trainable parameters. For each edge $e_k$ in the edge set $\mathcal{E}$, via the matrices $R_o$ and $R_i$, the feature vectors $b_o$, and $b_i$ of the two nodes at the extremities of $e_k$ are concatenated in a new edge message vector $B^k$. Each $B^k$ is then fed to a MLP with a single output neuron, and a sigmoid activation function, which bounds the scalar feature associated to each $e_k$ between 0 and 1. The resulting vector of scalar edge scores represents the learned weight associated with each edge.
                \newline
                Track segments with score close to 1 are to be considered more likely to be true, and therefore the information of the nodes at their extremities should be propagated with higher weight during message passing. The final stage of the GNN is an Edge Network, because the end goal of the pipeline is to get track segments' scores, and, by thresholding them, a classification of true and fake track segments. In this work the final threshold is always 0.5 .

            \subsubsection{Node Network}
            
                The Node Network (NN) block is responsible for the single-hop message passing in an event graph, where each node $j$ concatenates its own feature vector with the ones belonging to every other node with which it shares an edge, weighted by the edge score computed by the previous Edge Network block.
                As for the previous EN, the concatenated feature vector matrix is then fed row-by-row to a MLP.

\section{Quantum Graph Neural Networks}\label{S:QGNN}

    First introduced in~\cite{tuysuz2021hybrid}, quantum graph neural networks (QGNN) for particle tracking extend the GNN formalism into the domain of quantum computing by implementing parts of the architecture as parameterized quantum circuits (PQC). The core idea is to reproduce, and potentially enhance, the representational power of a classical multi-layer perceptron, used in standard GNNs, by alternating classical MLPs blocks with a single PQC as central component of both Edge and Node Networks.
    
    The research questions that this study, as the original one, sets itself to address are:
    \begin{itemize}
        \item[1:] \emph{Can a hybrid quantum graph neural network learn relationships in classical data from tracking in HEP as good as --or better-- than a classical GNN?}
        \item[2:] \emph{What constraints do current NISQ hardware and software frameworks impose on the tasks that quantum machine learning can realistically perform in the context of experimental high energy physics?}
        \item[3:] \emph{Can the original architecture in~\cite{tuysuz2021hybrid} be upgraded to better fit the particle tracking task?} 
    \end{itemize}

    To provide answers to these questions this project was developed in two phases, presented in the following sections. Phase I comprises the software development of a new, up-to-date version of the QGNN model, a study at increasing levels of collision pileup, and a test on quantum hardware. Phase II builds on the results and new evidences gathered in the previous phase, presenting the development of an upgraded QGNN architecture, together with a final characterization.

    \subsection{Phase I: development and pileup studies on the original architecture}
    \label{S:p1}

        \begin{figure*}
            \centering
            \includegraphics[width=0.5\linewidth]{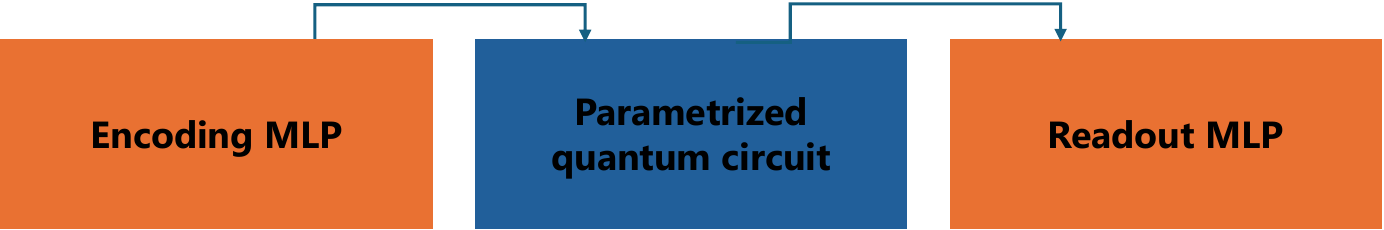}
            \caption{Structure of a generic hybrid sub-module in the QGNN, with encoding and readout layers to manage the dimensionality of the data, and a parametrized quantum circuit between classical networks. Arrows represent the data flow; no additional learnable parameters are present in-between different blocks, and the output of the previous block is directly taken as input to the next.}
            \label{fig:qgnn}
        \end{figure*}
    
        In the first phase of this project the original software of the QGNN in~\cite{tuysuz2021hybrid} was completely reimplemented and updated to use modern frameworks for both the classical machine learning components, and the integration with quantum circuits. The original study uses an integration of TensorFlow, TensorFlow Quantum~\cite{broughton2021tensorflowquantumsoftwareframework}, and Cirq~\cite{CirqDevelopers_2025}. Instead, two new versions have been developed: a first PyTorch - Qiskit~\cite{qiskit} implementation was initially tested, but due to the strong limitations in the Qiskit quantum machine learning interface \verb|TorchConnector| (such as the lack of GPU-accelerated quantum simulation and slow integration of backpropagation through the classical MLPs and simulated circuits), a second and final implementation was developed with Jax~\cite{jax2018github} and Flax~\cite{flax2020github} as the machine learning framework, and Pennylane~\cite{bergholm2022pennylaneautomaticdifferentiationhybrid} for the parametrized circuits. This final version, through a custom integration of the functionalities of Jax with the simulation backends offered by Pennylane, has been optimized to perform the training of all the QGNN models within the timescale of 24 hours, allowing tests that the previous implementations could not handle in reasonable time frames.

        In this first phase the underlying architecture is kept consistent with the original proposal. In particular the Phase I QGNN exploits the same structure of the GNN presented in the previous section, integrating the parametrized quantum circuits inside the Edge and Node Network blocks as depicted in Fig.~\ref{fig:qgnn}. While in a purely classical GNN the network itself is a fully connected deep neural network, in this QGNN the dense layers are alternated with parametrized quantum circuits, one for each Edge and Node Network.

        The encoding and readout layers adjust the number of features to be fed into or read out of the circuit. In this phase, tailored to the original architecture, the number of hidden neurons of the single layer classical networks is $n_{\text{hid}}=4$, to match the number of qubits of the quantum circuits. The circuit itself consists of three stages. The first stage is the \emph{information encoding circuit} (IEC), which is responsible for encoding classical features into a quantum state. In Phase I the QGNN uses angle encoding on 4 qubits: the classical features, output of the encoding MLP, are each associated to single-qubit Y-axis rotation gates. This encoding limits the depth of the IEC, but potentially requires a large number of qubits to encode bigger feature vectors. 

        \begin{figure}
            \centering
            \includegraphics[width=1\linewidth]{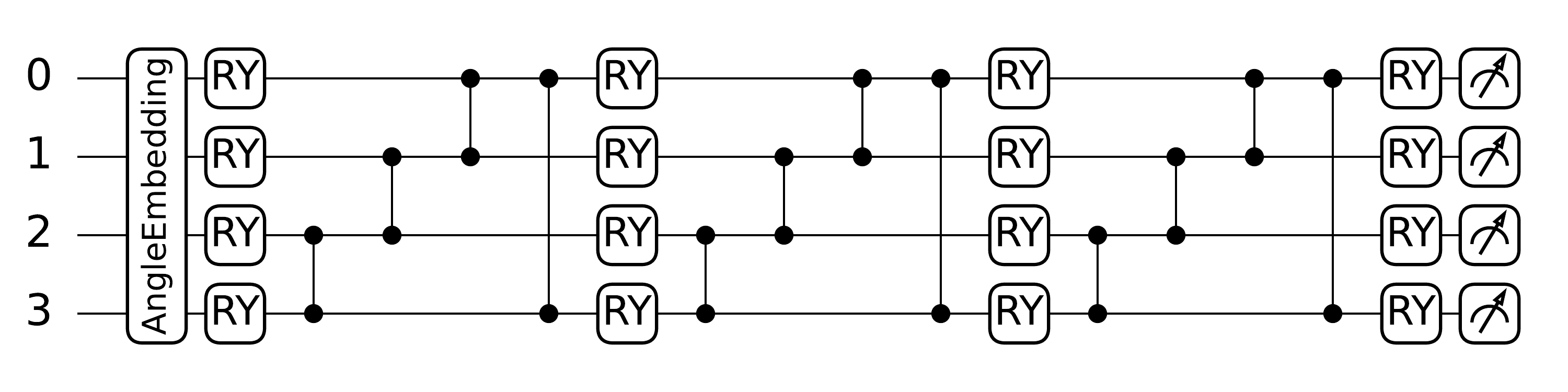}
            \caption{The 3-layers and 4-qubits parametrized quantum circuit used as foundation to the original hybrid classical-quantum layers of the QGNN. Information is encoded in the initial quantum state through angle encoding. $RY$ gates perform $Y$-axis rotations, each parametrized by a trainable angle $\theta_i$}
            \label{fig:circ10}
        \end{figure}
        
        The quantum state is directly fed to a parametrized quantum circuit with learnable rotation angles. The circuit, shown in Fig.~\ref{fig:circ10}, and characterized in~\cite{Sim_2019}, is chosen to reproduce the original study~\cite{tuysuz2021hybrid}, and is composed of 3 repeated layers.
        
        Finally, classical information is extracted through measurement gates, and fed to a readout network with a single output neuron in the case of Edge Networks, and 4 neurons for Node Networks.

\subsubsection{Training characterization}

        The characterization of the QGNN is performed through the creation of three new datasets, simulating increasing pileup values $\mu=50,100,150$, in addition to the original dataset at $\mu=200$.
        This is done through a random selection of the original $\sim 200$ primary interaction vertices.

\begin{figure*}
            \centering
            \includegraphics[width=0.91\linewidth]{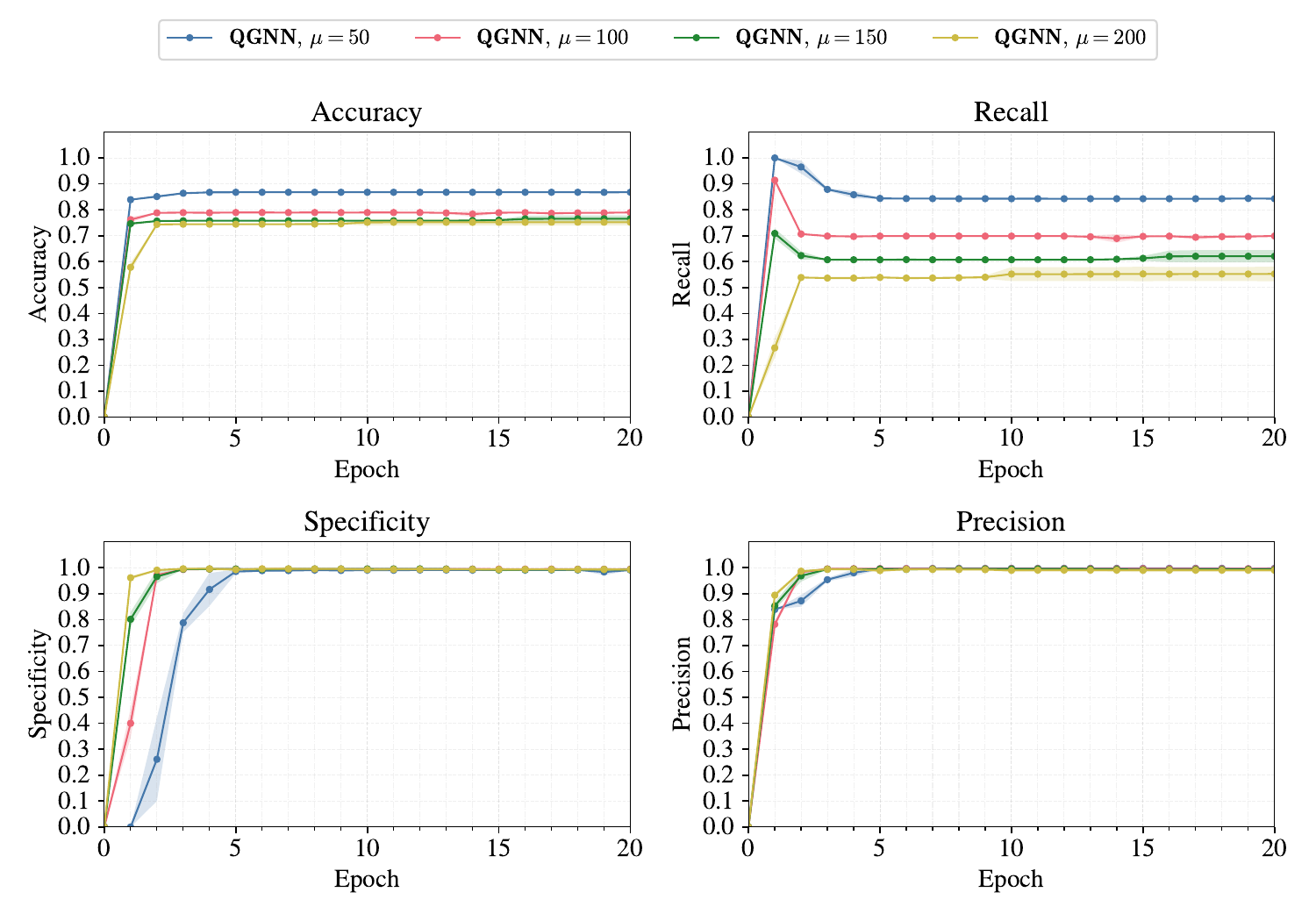}
            \caption{Original QGNN architecture accuracy, specificity, precision, and recall measured on the validation set, with four different dataset simulating different pileup values, $\mu=50,100,150,200$.}
            \label{fig:validation_metrics_original}
        \end{figure*}

        The same QGNN architecture is trained on the four datasets and the results are shown in Fig.~\ref{fig:validation_metrics_original} for the main validation metrics (refer to Appendix~\ref{A:train} for further characterization). The datasets are increasingly unbalanced for decreasing pileup, due to the constraint on using the same preprocessing for the four graph-building stages. As the pileup decreases, in fact, the geometrically available hits resulting in fake track segments also decrease.
        Error bars in the plots are obtained by k-folding, by splitting the training set in k equal parts (folds), and using $k-1$ of them as a new training set for each of the $k$ separate trainings. The remaining fold is, for each training, used as validation set. Averaging the k validation scores provides a reliable performance estimate. 
        During training, the quantum circuits are simulated on an ideal noiseless statevector simulator.
    
        The results show that the validation accuracy decreases with increasing pileup, as expected since the events become denser. The numerical value of the accuracy plateau for the $\mu=200$ dataset, corresponding to high luminosity events, is slightly dependent on the choice of the hyperparameters of the neural network, as the error estimation highlights. In this study the current set of hyperparameters, with learning rate 0.01, and training set consisting of 45 samples, is chosen so that one training for $\mu=200$ can be completed in less than one day. The development of the current software has been crucial to achieve this result, overcoming the limitations of the original implementation which is reported to train in the time-frame of a week~\cite{tuysuz2021hybrid}.
    
        The remaining metrics are particularly relevant; in fact, specificity and precision show that the different models are very good at recognizing fake track segments, while recall suggests that true track segments are often misclassified.

        These results highlight an unexpected behavior from the original QGNN. The model at this stage is not in fact learning to correctly classify true track segments: this can be inferred from the low value of the recall metric for $\mu=200$, which in turn provides an interpretation for the low accuracy score on the same dataset. Further characterization of this model on different backend simulators and quantum hardware is reported in Appendix~\ref{A:Inf}.

\subsection{Phase II: upgrades to the original QGNN model}
    \label{S:p2}

        Phase I of this project has revealed the practical limitations of the original QGNN architecture. The low accuracy plateau reached on the pileup $\mu=200$ dataset, which is consistent with the results in the original study~\cite{tuysuz2021hybrid} (small differences, in the order of the percent, can be linked to minor differences in the preprocessing step), indicate that the original QGNN architecture does not model the track segment classification with sufficient accuracy.

        At the start of this second phase the critical aspects motivating an upgrade can be grouped in three classes:
        \begin{itemize}
            \item \textbf{Dataset:} Particle tracking on a realistic dataset is extremely resource-demanding in the context of NISQ era quantum machine learning. The presence of classical MLPs can mask some of this complexity.
            \item \textbf{GNN:} The underlying classical GNN in the original proposal can be improved by taking into account advancements in classical GNN-based tracking pipelines such as those in Refs.~\cite{ExaTrkX:2021abe, Acorn:2024, Correia:2919388}.
            \item \textbf{QGNN:} The full hybrid pipeline is limited by bottlenecks in the classical-quantum interfaces.
        \end{itemize}
        
        \subsubsection{GNN upgrades}
            To answer the initial research questions in a meaningful way, the preprocessing cuts on the TrackML dataset are kept unchanged with respect to Phase I. The goal of the following upgrades is to improve both the training and the generalization performance on the $\mu=200$ dataset.

            The upgrades to the classical GNN involve two main aspects:
            \begin{itemize}
                \item The introduction of residual connections in the forward data flow.
                \item The tests of larger, more expressive MLPs, and the interplay between their outputs and the input features of the quantum circuits from the original architecture.
            \end{itemize}

            Residual connections have been shown to improve trainability and stability of deep message passing models, by explicitly preserving information across different layers~\cite{li2019deepgcns}. A residual block learns a correction function $\hat{R}(x)$ such that, denoting by $H^{(k)}$ the hidden representation at layer $k$ of a deep neural network:
            \begin{equation}
                H^{(k+1)}=H^{(k)}+\hat{R}\left(H^{(k)}\right).
            \end{equation}
            This has the combined effect of: (i) providing an explicit path so the network can default to the simple non-transformed branch when deeper transformations are unnecessary~\cite{he2016deep}; (ii) improving gradient flow because derivatives can propagate in the additive identity branch~\cite{he2016identity}; and (iii) smoothening the loss function landscape, enabling substantially deeper stacks without degradation~\cite{li2018visualizing}.
            These improvements, historically first introduced in deep feedforward neural networks, are beneficial also for graph neural networks~\cite{chen2020simple}. In fact, residual connections counteract feature oversmoothing, happening in the process of propagation, through the re-injection of pre-message-passing features. This way node representations do not lose significancy as depth grows.
            A similar, but complementary, method used with respect to the original network, and inspired by~\cite{ExaTrkX:2021abe} and~\cite{chen2020simple}, is an initial residual. Each node network output concatenates the original $H(0)$ with the current hidden state before the residual add; this preserves raw features that might otherwise be attenuated by repeated neighborhood mixing.

            These upgrades alone are not sufficient to improve the scores of the GNN, and need to be complemented by wider and deeper MLPs, which substitute the single layer and 4-neurons MLPs of the original architecture. In particular, referring to the studies by~\cite{ExaTrkX:2021abe} and~\cite{Acorn:2024}, a 2-hidden units MLP is introduced before the circuits, with 64 neurons per layer. Further tests with double the number of neurons have not shown significant advantage.

            The validation results of the new GNN architecture are shown, together with the results from next section, in Fig.~\ref{fig:validation_200_full}. All the metrics for the upgraded Classical GNN exceed 0.95, and suggest that the learning limitations observed in the previous model are effectively addressed in the classical network.

\subsubsection{QGNN upgrades}
        \label{S:QGNN_upg}

            \begin{figure}
                \centering
                \includegraphics[width=1\linewidth]{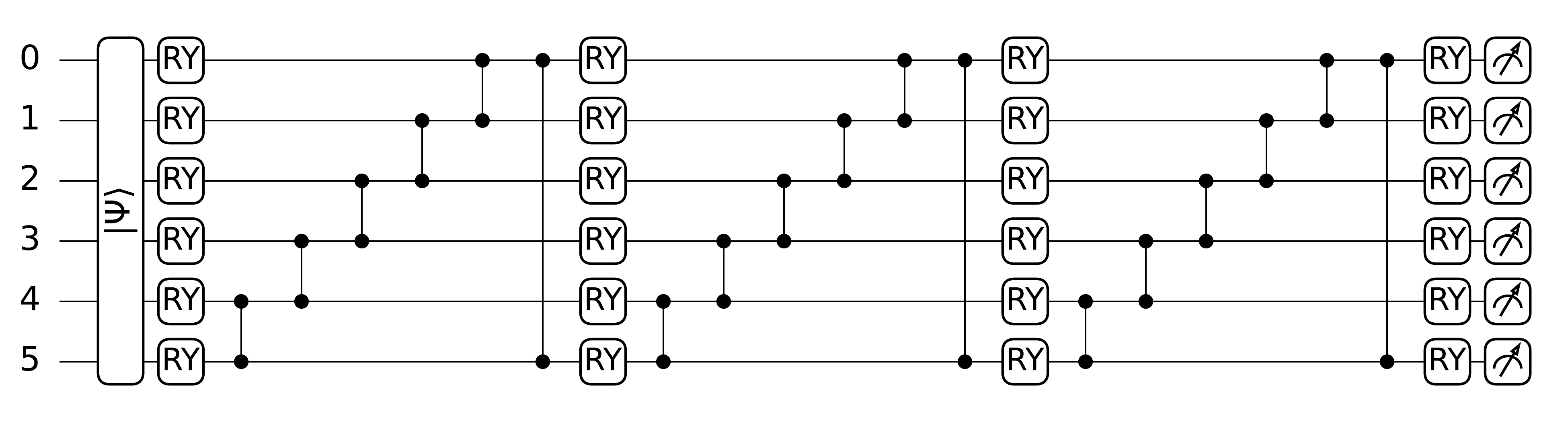}
                \caption{The 3-layers and 6-qubits parametric quantum circuit used in the upgrade. The 64 features from the classical MLPs are encoded in the quantum state $\ket{\psi}$ through amplitude encoding; 64 output amplitudes are then extracted from the circuit.}
                \label{fig:circ10_6_q_amplitude}
            \end{figure}         

            \begin{figure*}[]
                \centering
                \includegraphics[width=0.91\linewidth]{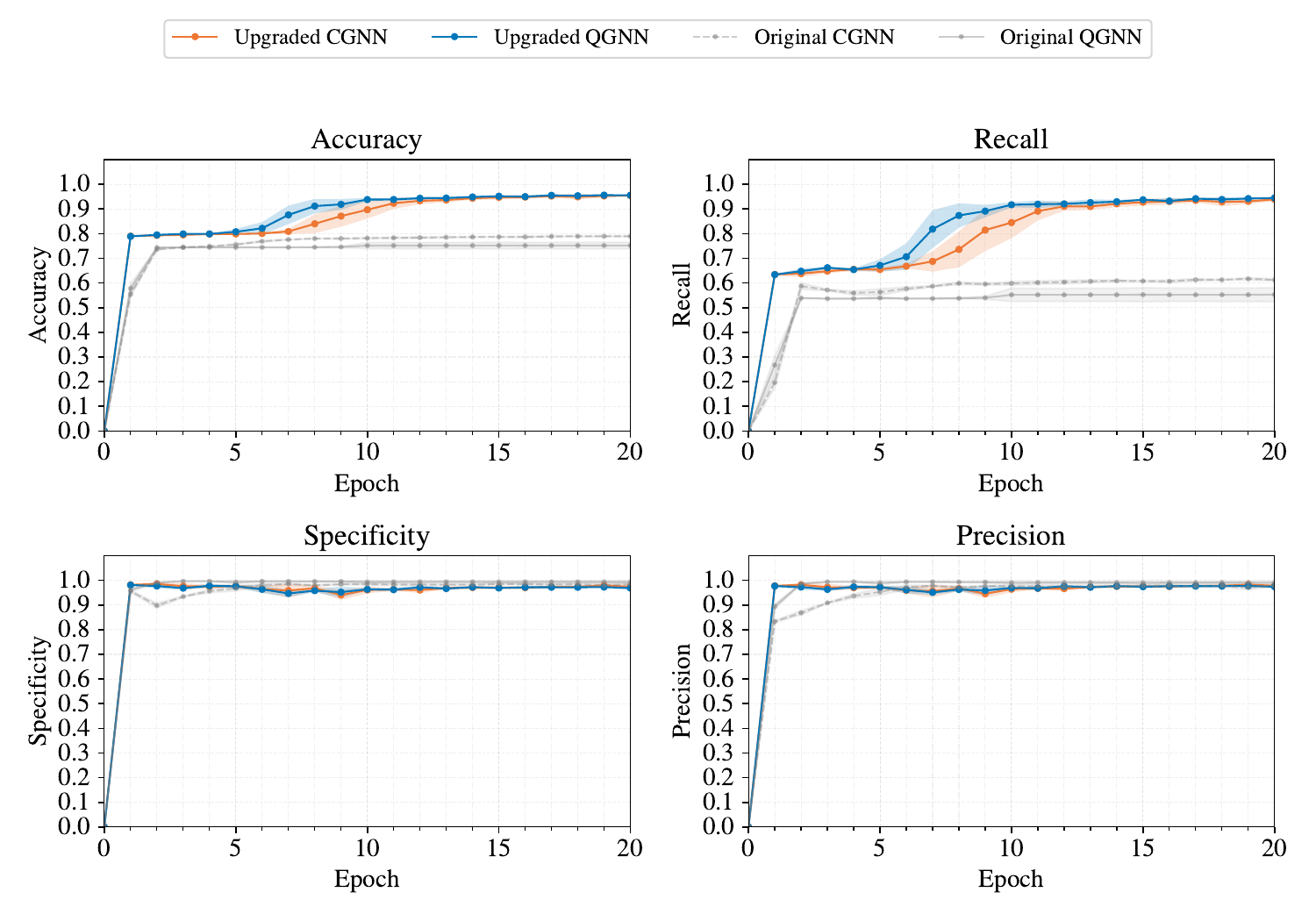}
                \caption{Comparison of the learning metrics on the $\mu=200$ validation set. Blue lines represent data from the upgraded Quantum-GNN architecture, with [64,64] classical MLPs and 6-qubits quantum circuit with amplitude encoding. Orange lines correspond to data from the upgraded Classical-GNN, with [64,64] MLPs. Grey lines represent the original pre-upgrade architecture scores.}
                \label{fig:validation_200_full}
            \end{figure*}

            \begin{table*}[]\label{tab:final_results}
            \centering
            \renewcommand{\arraystretch}{1.25}
\begin{tabular*}{\textwidth}{l @{\extracolsep{\fill}} c c c c}
            \toprule
                  \textbf{Model}
                & {\textbf{Accuracy}}
                & {\textbf{Precision}}
                & {\textbf{Recall}}
                & {\textbf{Specificity}} \\
            
            \midrule         
            
            Upgraded QGNN
                & $0.960 \pm 0.003$ & $0.981 \pm 0.003$ & $0.946 \pm 0.005$ & $0.977 \pm 0.004$ \\
            
            Upgraded CGNN
                & $0.964 \pm 0.002$ & $0.981 \pm 0.005$ & $0.952 \pm 0.003$ & $0.977 \pm 0.007$ \\  
            \midrule 
            Original QGNN
                & $0.752 \pm 0.013$ & $0.990 \pm 0.007$ & $0.55 \pm 0.03$ & $0.993 \pm 0.006$ \\
            
            Original CGNN
                & $0.790 \pm 0.003$ & $0.981 \pm 0.004$ & $0.612 \pm 0.004$ & $0.987 \pm 0.003$ \\
            \bottomrule
            \end{tabular*}
            \caption{%
                Final-epoch $k$-fold cross-validation performance
                of the upgraded and original reference models at $\mu = 200$.
            }
            \end{table*}
        
            Increasing the size of the MLPs of the Edge and Node Network blocks, detailed in the previous sections, leads to a nontrivial interplay with the quantum circuits in the full QGNN architecture. In Phase I, in-fact, 4-qubits quantum circuits with angle encoding of the classical information were used, in accordance with the limited 4-neurons dimension of the MLPs. After the upgrade, however, the new MLPs produce 64-dimensional feature vectors.

            A first characterization has been carried out directly on 4-qubits circuits, with an encoding MLP with hidden units sized (64,64,4), where the final 4 neurons of this sequence are set to correctly output 4 features for the quantum angle encoding. This results in the same plateaus of Phase I, indicating that the feature reduction, from 64 to 4, is discarding too much information. The pipeline trains, but again fails to learn the task, in this case due to the limited dimension of the quantum circuits.

            To circumvent this problem two different paths have been considered:
            \begin{itemize}
                \item[1:] Increase the number of qubits while keeping an angle encoding IEC.
                \item[2:] Switch to an amplitude encoding scheme.
            \end{itemize}

            The first strategy is conceptually straightforward but severely constrained by the cost of simulating a QGNN built with such a technology. Tests have been performed with up to 12 qubits, which have proved to be still not enough to avoid the information loss starting from the 64 features of the MLPs. This constraint comes from two counteracting effects: simulating a single circuit of 12 qubits is extremely fast, but this QGNN operates at edge-level, therefore, for each event graph, each node network creates $N_\mathcal{V}$ quantum circuits. Similarly each edge network creates $N_\mathcal{E}$ circuits. The number of message passing operations is fixed to $N_{\text{iter}}=3$ and this amounts to 7 total ENs and NNs, per graph. The total number of circuits to be simulated for graphs sized $N_\mathcal{V}\simeq5\cdot10^3$, and $N_\mathcal{E}\simeq8\cdot10^3$ is of order $N_{\text{tot}}\simeq1\cdot 10^5$ per event. Caching intermediate results for the backpropagation becomes easily unfeasible for circuits with more than 12 qubits. Similarly, avoiding the majority of the caching results in a very rapid rise in training times, of the order of weeks for meaningfully sized datasets.        
            
            For these reasons the upgraded architecture adopts amplitude encoding of the classical features, in the $2^{N_{\text{qubits}}}$ complex amplitudes of a system of $N_{\text{qubits}}$. The quantum circuit (Fig.~\ref{fig:circ10_6_q_amplitude}) is an adaptation of the original circuit in Fig.~\ref{fig:circ10}, now extended to $N_{\text{qubits}}=6$, in accordance with the 64-feature vectors of the MLPs.

            The results of the tests on this upgraded QGNN architecture on the validation set are shown in Fig.~\ref{fig:validation_200_full}, and in Tab.\ref{tab:final_results} . Accuracy and recall show drastic improvements with respect to the same metrics on the Phase I QGNN of Fig.~\ref{fig:validation_metrics_original} for the $\mu=200$ dataset. The upgraded QGNN is able to match, within uncertainties, the performance of the upgraded classical GNN. Comparisons with the original classical and quantum architectures show an overall improvement of both the upgraded classical and upgraded quantum models, both in terms of convergence within the two models, and in the absolute scores.
            
            Additionally, Fig.~\ref{fig:validation_200_full} shows that the upgraded QGNN displays a noticeable earlier convergence with respect to the upgraded CGNN. This behavior emerges similarly in the training curves reported in Appendix~\ref{A:train}.
            \newline
            The total number of classical trainable parameters is the same for both the architectures, full classical and hybrid quantum, and equal to $N_{\text{classical}}=29505$. The total number of trainable rotation angles in the quantum circuits of the QGNN is $N_{\text{quantum}}=44$. This pronounced mismatch is an inevitable feature of a NISQ hybrid QGNN designed for edge classification on graphs from realistic events at $\mu=200$, because much of the learning in this regime has to be offloaded to the more scalable classical MLPs.
            
            Nonetheless, the difference in convergence introduced by the presence of a quantum circuit, however limited in depth, suggests that quantum neural networks, or classical architectures inspired by them, may provide useful effects such as regularization or alternative inductive biases. Further studies are required to understand how dataset and preprocessing dependent is this effect.

\section{Conclusions}

    The research presented in this paper illustrates the delicate interplay between deep neural networks and parametrized quantum circuits, in the context of a quantum graph neural network for edge classification. This integration is particularly challenging for a complex task such as particle tracking at high pileup.

    The characterization, through increasing pileup studies, performed in the first phase of this project, has led to a better understanding of the limitations of the original proposals regarding QGNNs for edge classification. In particular the software development carried out during this first phase has built a solid foundation for the upgrades of the subsequent phase.

    In the second phase of the project an upgraded version of the QGNN is characterized. The results show a clear improvement with respect to the metrics of the original QGNN, and demonstrate that the upgraded hybrid model can match, within uncertainties, the performance of a classical GNN on the same task.

    The results also indicate that the underlying classical part of the architecture is responsible for the majority of the learning. This is an expected behavior for a NISQ architecture used in a real-world task. There are, however, indications of a possible regularization effect introduced by the quantum circuits during the learning phase, which results in a faster convergence of the QGNN, and motivates further research on hybrid GNNs.
\begin{acknowledgments}
    The research leading to these results has received funding from the European Union - NextGenerationEU through the Italian Ministry of University and Research under PNRR - M4C2 - I1.4 Project CN00000013 “National Centre for HPC, Big Data and Quantum Computing”.
\end{acknowledgments}



\bibliographystyle{apsrev4-1}
\bibliography{bibliography}

\appendix

\onecolumngrid
\newpage

\section{TrackML dataset and preprocessing}\label{A:TrackML}
    
    \begin{figure}[h]
        \centering
        \includegraphics[width=0.4\linewidth]{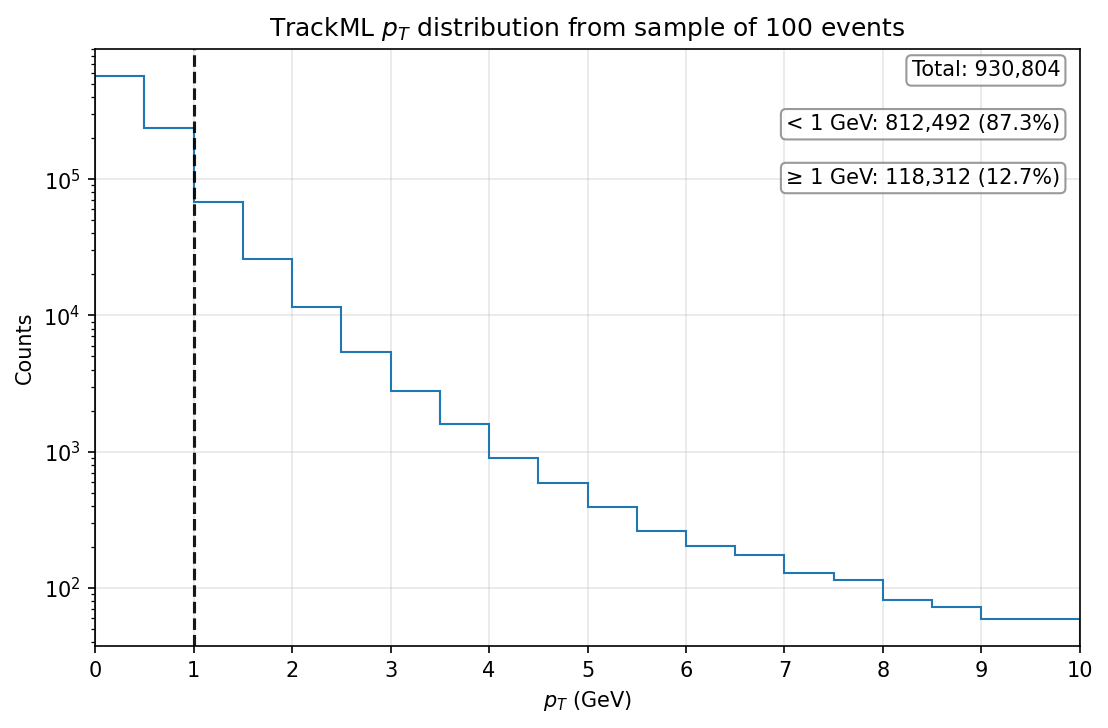}
        \caption{Distribution of transverse momentum $p_T$ in the 100-events sample dataset from~\cite{trackml-particle-identification}. The dashed black line represent the cut enforced at 1GeV.}
        \label{fig:distr_pt}
    \end{figure}

    TrackML~\cite{trackml-particle-identification} is a simulated dataset of collision events detected by the synthetic tracker shown in Fig.~\ref{fig:TrackML_detector}, and composed of three sub-modules:
    \begin{itemize}
        \item a silicon pixel detector with spatial segmentation $50\mu m \times 50\mu m$;
        \item a silicon short strip tracker with spatial segmentation $80\mu m \times 1200\mu m$;
        \item a silicon long strip tracker with spatial segmentation $0.12 mm \times 10.8 mm$.
    \end{itemize}

    The dataset is based on a Pythia 8 event generator, with, for each event, a hard QCD interaction producing a $t\overline{t}$ pair, and 200 additional soft QCD interactions, simulating the average predicted pileup of HL-LHC. The detector simulation models an ATLAS-like inhomogeneous magnetic field, and the propagation of the particles through the trackers' layers is obtained by an ACTS-based~\cite{ai2022common} simulation, accounting for different material interactions (multiple scattering, energy loss, and hadronic interactions). The full simulated dataset consists of particles with transverse momentum $p_T\ge 150$MeV. This work enforces a cut in the track transverse momentum $p_T$ of $1GeV$ to constrain the considered particles to the reasonably high momentum ones (Fig.~\ref{fig:distr_pt}).
    
\section{Graph Neural Networks formalism}\label{A:GNN}

This appendix serves as a compact reference to the mathematical formalism underlying the classical graph neural network architecture. The formalism is kept consistent with the original study~\cite{tuysuz2021hybrid}.

The input data structures are the feature matrix $X\in\mathbb{R}^{N_{\mathcal{V}}\times3}$, the incident matrices $R_i,\ R_o\in {\{0,1\}}^{N_{\mathcal{V}}\times N_{\mathcal{E}}}$ and the target vector $y\in{\{0,1\}}^{N_{\mathcal{E}}}$ with the ground truth about true and fake edges.

Within the \emph{InputNet} each row of $X$, representing the physical coordinates of a particle hit, is fed to a feedforward network which performs feature expansion, producing a new matrix $X^\prime$ with dimensions $N_{\mathcal{V}}\times d_{\textbf{hidden}}$, where $d_{\textbf{hidden}}$ is the dimension of the expanded features.

The new, expanded node features $X^\prime$ are then used inside the \emph{Edge Network} blocks to build a feature vector for each edge of the graph, according to the connectivity expressed by $R_o$ and $R_i$. For each edge $e_k$ in the edge set $\mathcal{E}$, the matrices $R_o$ and $R_i$ are used to build the feature vectors $b_o$, and $b_i$ of the two nodes at the extremities of $e_k$, which are then concatenated in a new edge message vector $B^k$:
    \begin{equation}
        B^k = [b_o^k\oplus b_i^k]\ ,
    \end{equation}
    where
    \begin{equation}
        b_o^k=\sum_{j=1}^{N_\mathcal{V}} R_o^{j k} x^\prime_j\ ,
    \end{equation}
    and
    \begin{equation}
        b_i^k=\sum_{j=1}^{N_\mathcal{V}} R_i^{j k} x^\prime_j.
    \end{equation}
Each $B^k$ is then fed to a dense network with one output neuron, and a sigmoid activation function, which bounds the scalar feature associated to each $e_k$ between 0 and 1. To avoid introducing unnecessary notation, from now on $e_k$ will refer both to the label of the k-th edge of the graph, and to its scalar weight. The resulting vector of scalar edge scores represents the learned weight associated with each edge.

The edge scores are used within the subsequent \emph{Node Network} block to update the feature vectors of each graph node by concatenating its old representation $x^\prime$ to the feature vectors of the incoming and outgoing edges $b_i^k$ and $b_o^k$, weighted by their edge scores $e_k$. Formally:

    \begin{equation}
        x_j^{\prime\prime} = [x_{j, \text {outgoing}}^{\prime\prime} \oplus x_{j, \text {incoming}}^{\prime\prime} \oplus x_j^\prime]
    \end{equation}
    \begin{equation}
        x_{j, \text {incoming}}^{\prime\prime}=\sum_{k=1}^{N_\mathcal{E}} e_k R_i^{j k} b_o^k
    \end{equation}
    \begin{equation}
        x_{j, \text {outgoing}}^{\prime\prime}=\sum_{k=1}^{N_\mathcal{E}} e_k R_o^{j k} b_i^k
    \end{equation}

As for the previous EN, the concatenated feature vector matrix $X^{\prime\prime}$ is then fed row-by-row to an MLP with output dimension $d_{\textbf{hidden}}$ equal to the number of expanded features of the Input network.

\section{Phase I inference results}\label{A:Inf}
        
        \begin{figure}[h]
            \centering
            \includegraphics[width=0.5\linewidth]{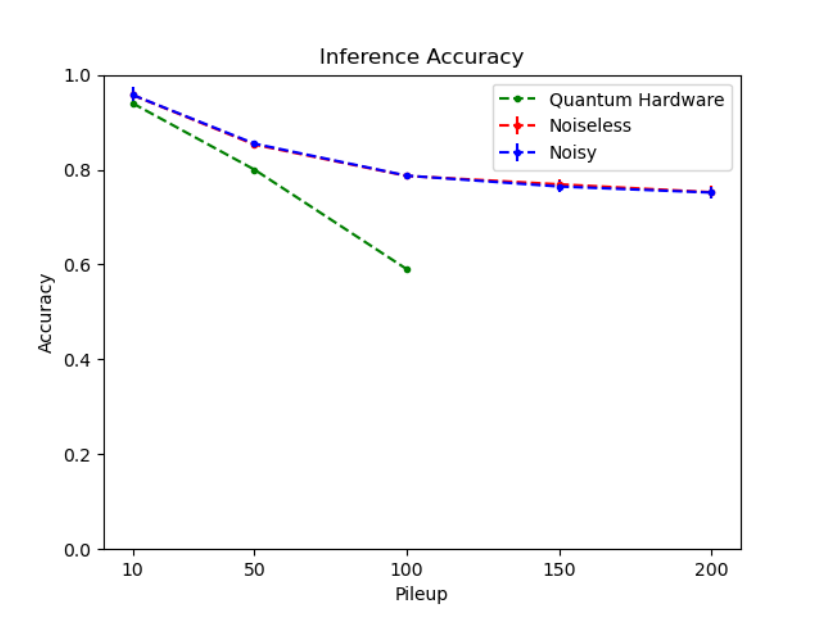}
            \caption{Inference accuracy for different pileup datasets. The red line corresponds to the noiseless PennyLane statevector simulator; the blue line to the noisy Qiskit Aer simulator backend, with a noise model imported from the IBM Osaka device; the green line to the real remote IBM Osaka quantum hardware, accessed through PennyLane’s interface.}
            \label{fig:inf}
        \end{figure}
    
        Another goal of this work is the characterization of the trained QGNN models on different backends, both simulated and real quantum hardware \footnotemark.
        \footnotetext[\value{footnote}]{
            \makebox[0pt][l]{%
                \parbox[t]{\textwidth}{%
                This work was supported by the PNRR MUR project ICSC under grant CN00000013-ICSC. Access to the IBM Quantum Services was obtained through the IBM Quantum Hub at CERN under the CERN-INFN agreement contract KR5386/IT.
                }%
            }%
        }
        Fig.~\ref{fig:inf} shows the accuracy with increasing pileup values. The results are compatible with the validation accuracy of Fig.~\ref{fig:validation_metrics_original}. For the simulated backends, tests rely on Pennylane's support for local IBM Qiskit Aer quantum simulators. The two lines representing noiseless and noisy simulated backends overlap almost completely. This behavior is compatible with the fact that the quantum circuits are limited in depth and therefore not very complex and subject to non-mitigated noise.

        In the first quarter of 2024, thanks to access to IBM's superconducting quantum hardware, the pre-trained models have been tested on the 127-qubit IBM Osaka machine up to pileup 100. The dataset sizes that could be tested are very limited (less than 10 events), and for this reason this result has to be considered only in its qualitative nature. The limitations in the tests are related to the high queuing times for submitting quantum jobs and the limited access to quantum processing time. The divergence between the green line and the simulated results can be likely attributed to two reasons: (i) the limited statistics due to the small dimension of the test dataset that can be reasonably tested on current quantum hardware, and the consequent inability to perform k-folding; (ii)~the sub-optimal IBM remote hardware support offered by Pennylane at the time of testing \footnotemark. 
        \footnotetext[\value{footnote}]{
            \makebox[0pt][l]{%
                \parbox[t]{\textwidth}{%
                \rule{0pt}{1.5em}%
                This issue has been dealt with in the meantime, but access to quantum hardware was no longer available for this project.
                }%
            }%
        }

\section{Phase I and II training results}\label{A:train}

\begin{figure}[h]
            \centering
            \includegraphics[width=0.90\linewidth]{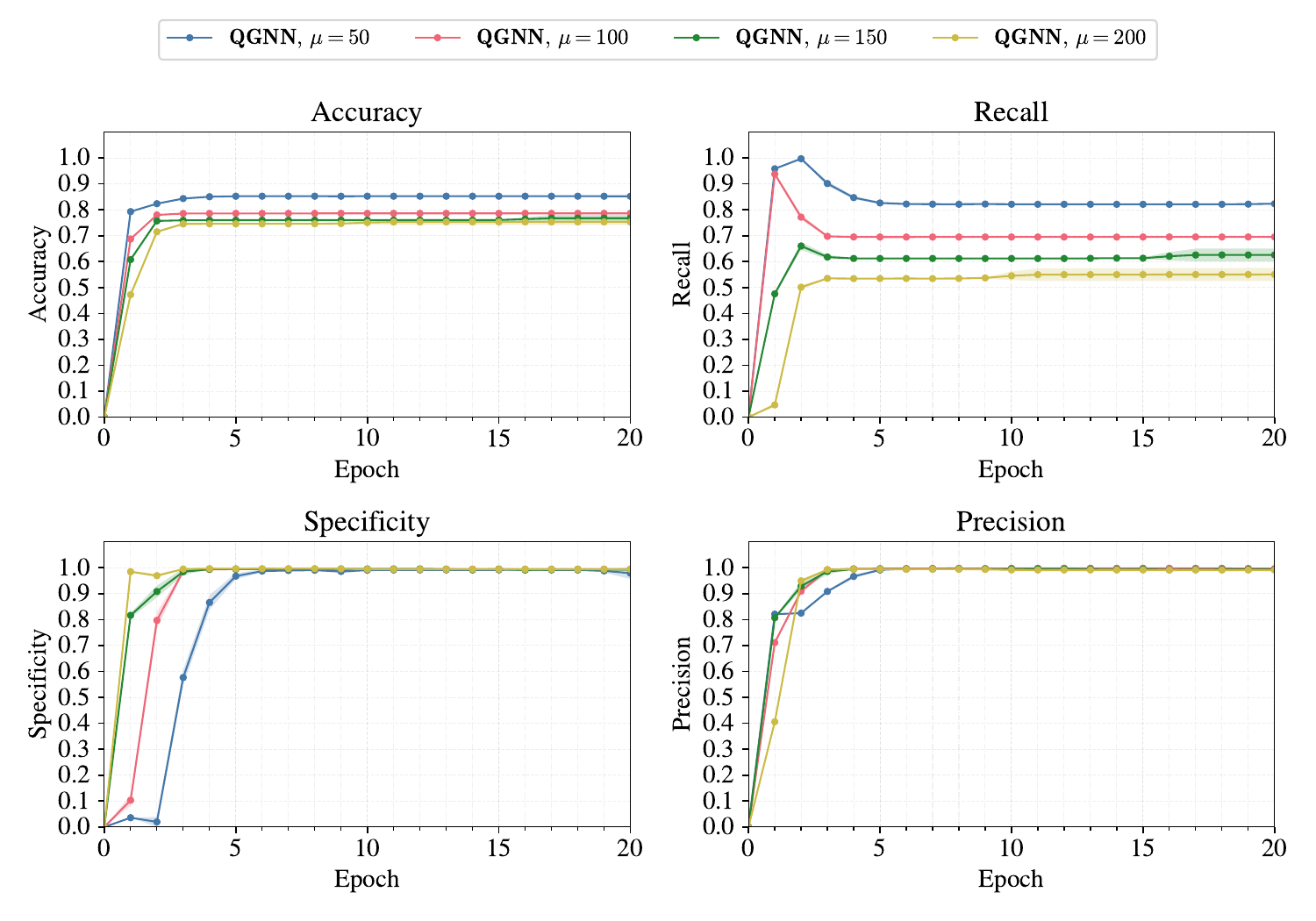}
            \caption{Original QGNN architecture accuracy, specificity, precision, and recall measured on the training set, with four different dataset simulating different pileup values, $\mu=50,100,150,200$.}
            \label{fig:training_metrics_original}
        \end{figure}

\begin{figure}[h]
            \centering
            \includegraphics[width=0.90\linewidth]{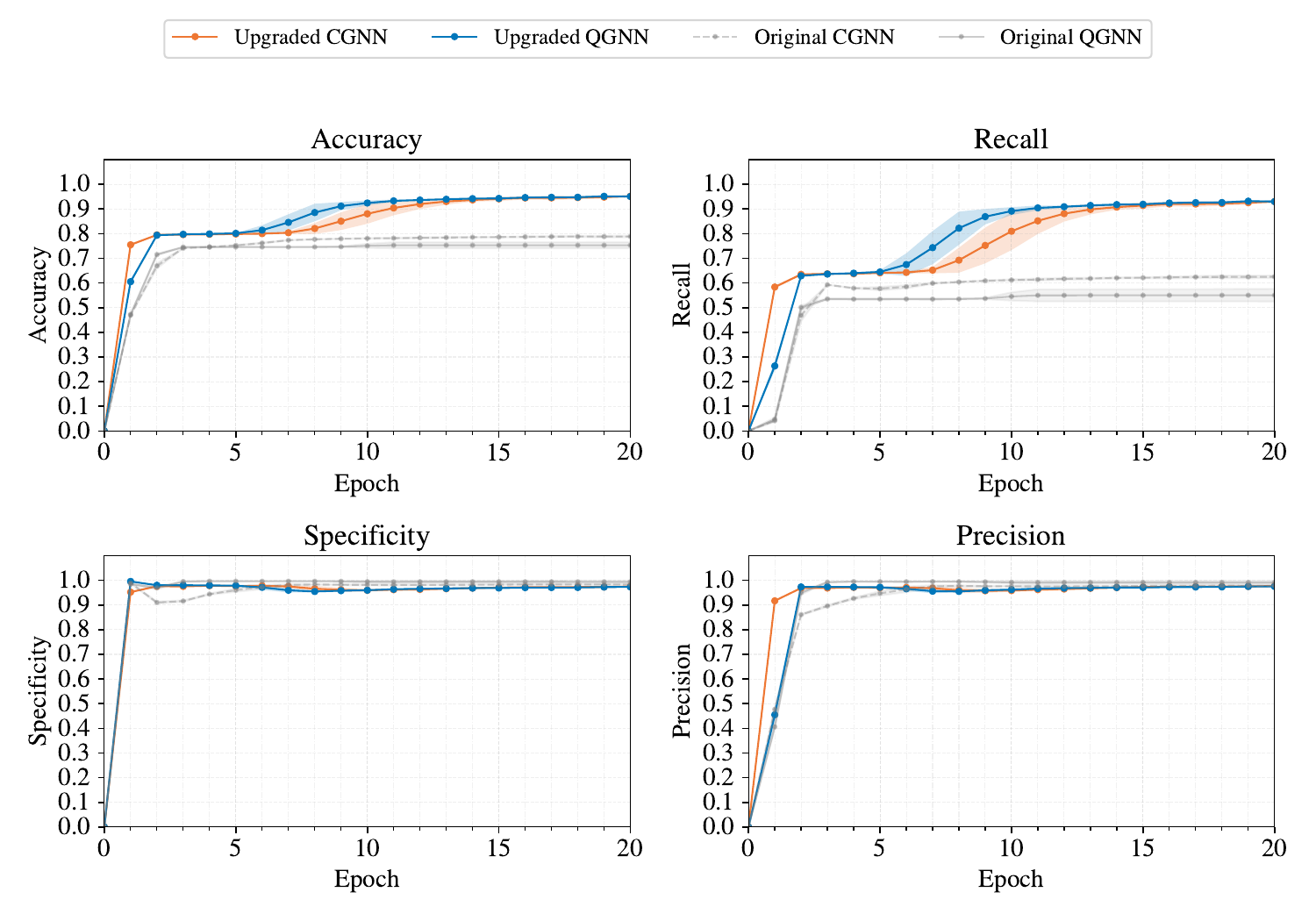}
            \caption{Comparison of the learning metrics on the $\mu=200$ training set. Blue lines represent data from the upgraded Quantum-GNN architecture, with [64,64] classical MLPs and 6-qubits quantum circuit with amplitude encoding. Orange lines correspond to data from the upgraded Classical-GNN, with [64,64] MLPs. Grey lines correspond to the original, pre-upgrade architectures.}
            \label{fig:training_metrics_full}
        \end{figure}

        Figures~\ref{fig:training_metrics_original} and~\ref{fig:training_metrics_full} report the training curves for, respectively, the original QGNN architecture at increasing pileup values, and the upgraded QGNN and CGNN, corresponding to the validation curves of Fig.~\ref{fig:validation_metrics_original}, and Fig.~\ref{fig:validation_200_full}.

\section{Phase II parallel encoding}\label{A:par}

        \begin{figure}[h]
            \centering
            \includegraphics[width=0.7\linewidth]{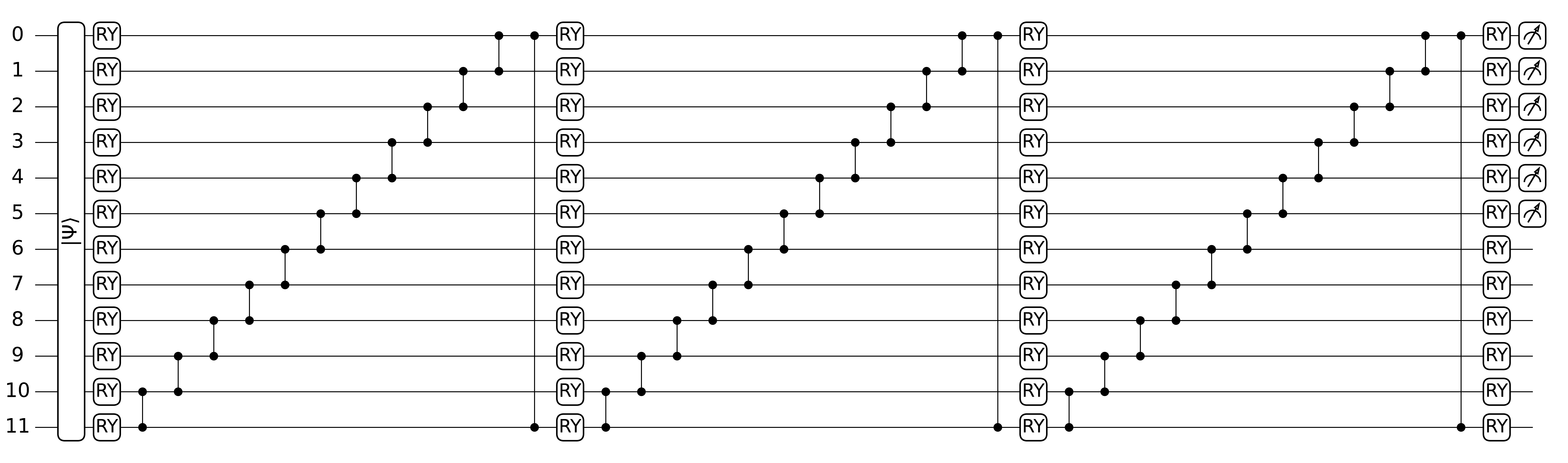}
            \caption{Extended 12 qubits circuit with $\ket{\psi} = \ket{\psi_{\text{in}}}\otimes\ket{\psi_{\text{in}}}$ tested within the upgraded QGNN architecture for preliminary studies on expressivity enhancements.}
            \label{fig:12q_circ}
        \end{figure}

        Further studies have been performed on the possible enhancements to the expressivity of the parametrized quantum circuits. As detailed in Section~\ref{S:QGNN_upg}, tests with data reupload techniques~\cite{P_rez_Salinas_2020, Schuld_2021} through angle embedding are not possible because of the limitations introduced by this kind of information encoding in terms of qubit number.

        To circumvent this problem, a repeated parallel amplitude encoding is studied, doubling the number of qubits to perform:
        \begin{equation}
            \ket{\psi} = \ket{\psi_{\text{in}}}\otimes\ket{\psi_{\text{in}}},
        \end{equation}
        where $\ket{\psi_{\text{in}}}$ is the 64-complex amplitudes state, input of the circuit used in the upgraded QGNN. The total number of qubits is therefore 12, with measure only on the first 6 qubits, to still sample 64-feature vectors. The adapted circuit is shown in Fig.~\ref{fig:12q_circ}.
        
\begin{figure}[]
            \centering
            \includegraphics[width=0.90\linewidth]{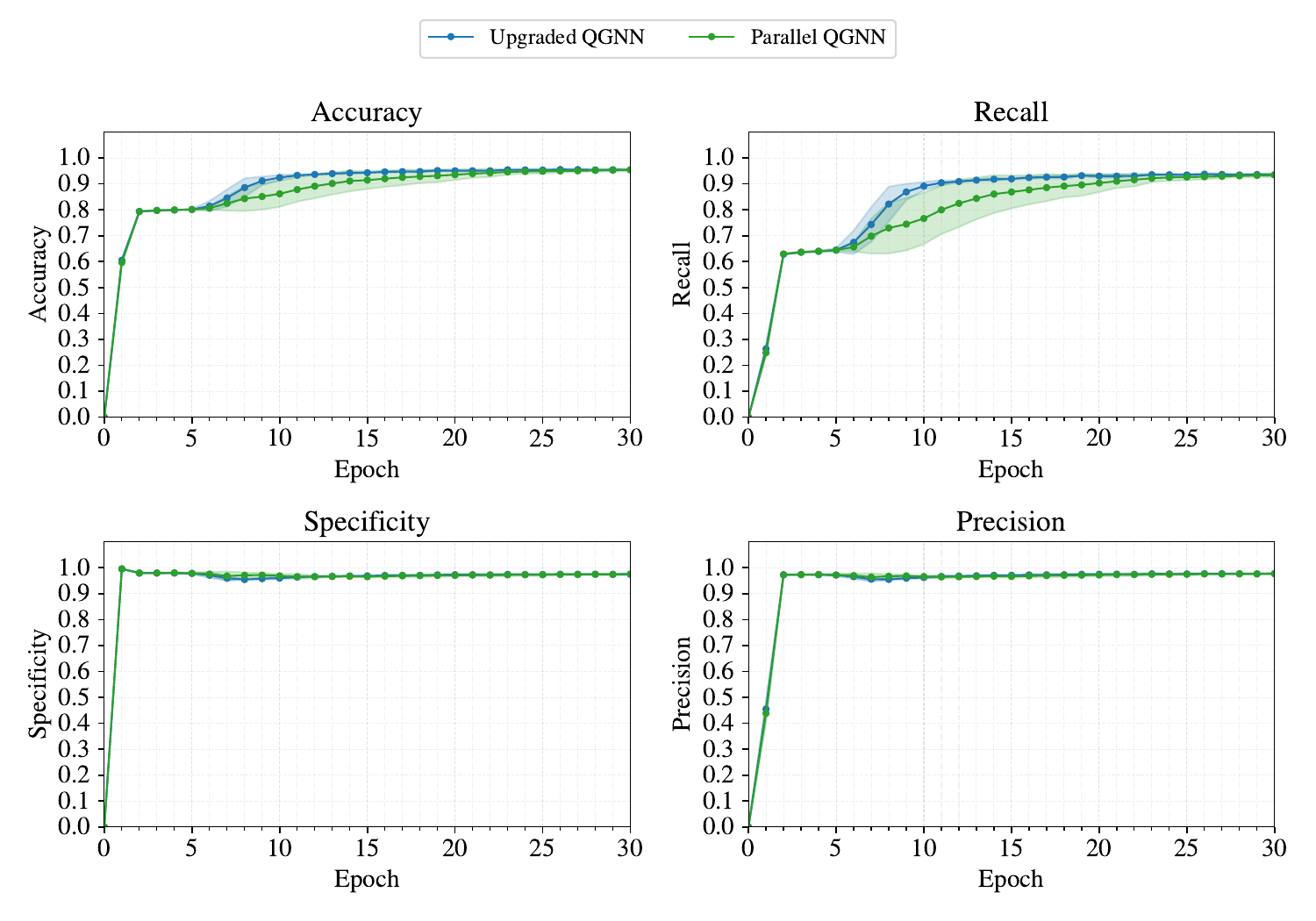}
            \caption{Comparison of the training metrics between the Upgraded QGNN and the same architecture with the circuit in Fig.~\ref{fig:12q_circ}, labeled Parallel QGNN.}
            \label{fig:training_metrics_reup_adapted}
        \end{figure}

\begin{figure}[]
            \centering
            \includegraphics[width=0.90\linewidth]{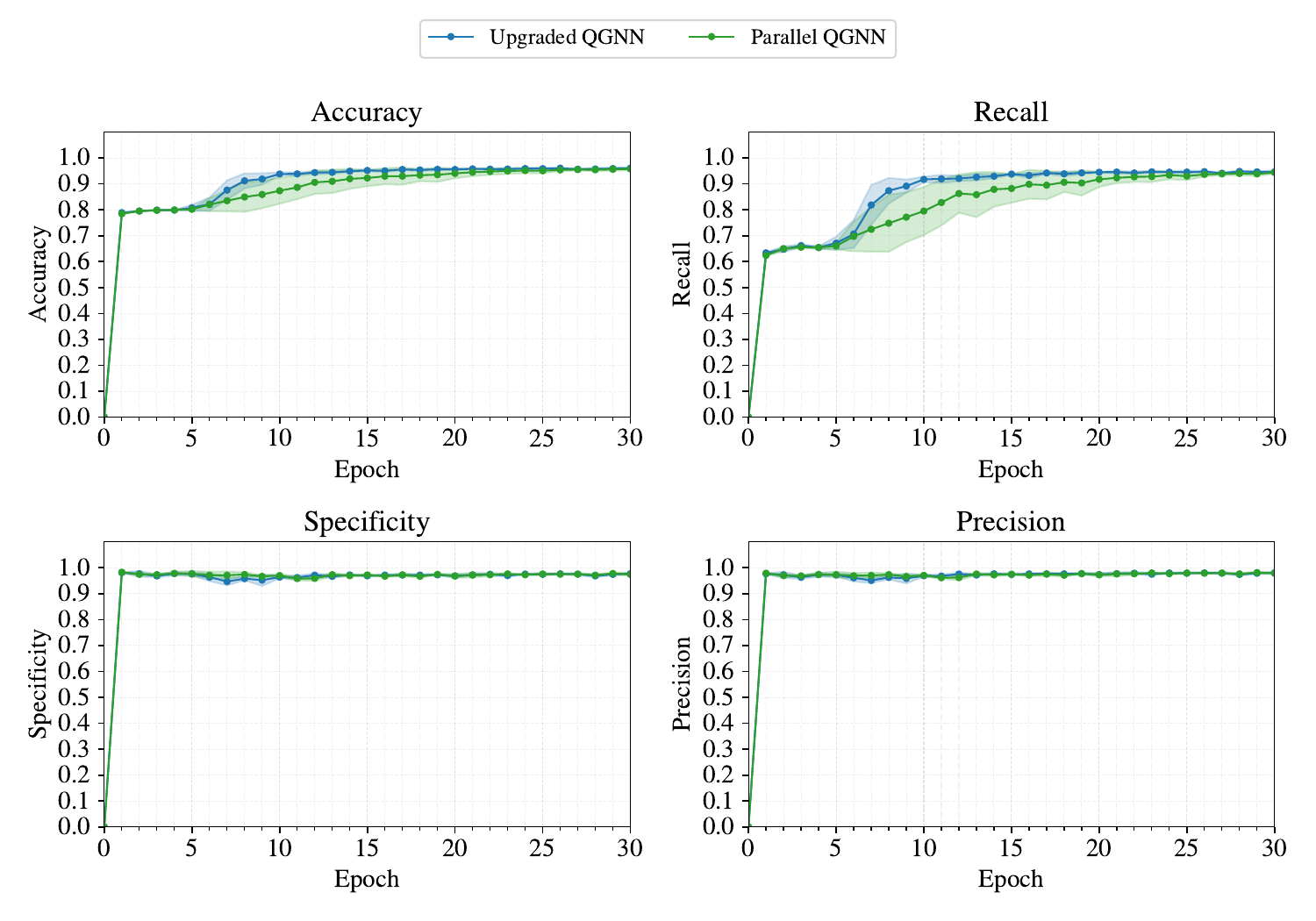}
            \caption{Comparison of the validation metrics between the Upgraded QGNN and the same architecture with the circuit in Fig.~\ref{fig:12q_circ}, labeled Parallel QGNN.}
            \label{fig:validation_metrics_reup_adapted}
        \end{figure}

        Fig.~\ref{fig:training_metrics_reup_adapted} and Fig.~\ref{fig:validation_metrics_reup_adapted} show an assessment of the training and validation metrics with this extended circuit. The new circuit introduces bigger variance in the training process, and on validation curves, but does not amount, at least in this preliminary assessment, to either higher metrics, or faster convergence with respect to the single-state amplitude embedding circuit.
        
        The full training process for these 12-qubits models, caching intermediate results to speed up the backpropagation, amounts during training to an average memory occupancy of $\approx300GB$, and are completed, for each of the repeated trainings, in $\approx60$hours on a modern Nvidia GH200 ~\cite{nvidia2023gh200}.


\end{document}